\documentclass[conference]{IEEEtran}
\IEEEoverridecommandlockouts
\usepackage{cite}
\usepackage{amsmath,amssymb,amsfonts}
\usepackage{graphicx}
\usepackage{textcomp}
\usepackage{xcolor}
\usepackage{algorithm}
\usepackage{algpseudocode}
\def\BibTeX{{\rm B\kern-.05em{\sc i\kern-.025em b}\kern-.08em
    T\kern-.1667em\lower.7ex\hbox{E}\kern-.125emX}}
\begin{document}

\DeclareRobustCommand*{\IEEEauthorrefmark}[1]{%
    \raisebox{0pt}[0pt][0pt]{\textsuperscript{\footnotesize\ensuremath{#1}}}}

\newcommand{\todo}[1]{\textbf{\textcolor{red}{@todo: #1}}}

\title{Reverse engineering the brain input: Network control theory to identify cognitive task-related control nodes\\

\thanks{This work was funded in part by the National Key R\&D Program of China (2021YFF1200804), Shenzhen Science and Technology Innovation Committee (2022410129, KCXFZ2020122117340001).}

}
\author{
\IEEEauthorblockN{
Zhichao Liang\IEEEauthorrefmark{1},
Yinuo Zhang\IEEEauthorrefmark{1},
Jushen Wu\IEEEauthorrefmark{1}, and
Quanying Liu \IEEEauthorrefmark{1}$^{*}$}
\IEEEauthorblockA{\IEEEauthorrefmark{1}Department of Biomedical Engineering, Southern University of Science and Technology, Shenzhen, 518055, China}
\IEEEauthorblockA{$*$ Corresponding Author: Quanying Liu \quad Email: liuqy@sustech.edu.cn}}

%

\maketitle
\begin{abstract}
The human brain receives complex inputs when performing cognitive tasks, which range from external inputs via the senses to internal inputs from other brain regions. However, the explicit inputs to the brain during a cognitive task remain unclear.
Here, we present an input identification framework for reverse engineering the control nodes and the corresponding inputs to the brain. The framework is verified with synthetic data generated by a predefined linear system, indicating it can robustly reconstruct data and recover the inputs.
Then we apply the framework to the real motor-task fMRI data from 200 human subjects.
Our results show that the model with sparse inputs can reconstruct neural dynamics in motor tasks ($EV=0.779$) and the identified 28 control nodes largely overlap with the motor system. 
Underpinned by network control theory, our framework offers a general tool for understanding brain inputs.
\end{abstract}


\section{Introduction}
When performing a complex cognitive task, multiple brain regions are engaged in sequence, for gathering sensory inputs, reasoning, making decisions, and executing actions\cite{pessoa2023entangled,zhang2020eeg}.
Our brain receives complex task inputs and hierarchically coordinates these task-related regions, from low-level sensory regions to high-level associative regions\cite{ito2020cortical,wang2021flexible,deco2021revisiting,liu2017detecting}. 
Uncovering which brain areas receive task-relevant inputs and how these inputs are sequentially transmitted to other brain areas are core scientific questions in neuroscience. 
These motivations for exploring task inputs to the brain system pose new technical challenges.

In neuroscience, task-related brain regions are typically identified through statistical analysis of functional magnetic resonance imaging (fMRI)\cite{grady2021influence}, including univariate analysis (e.g., general linear model (GLM) and t-test) and multivariate analysis (e.g., repeated measures ANOVA). Specifically, GLM regresses each voxel of fMRI data with task-related variables (e.g., task onset, reaction time, or behavioral performance). The regions that demonstrate statistical significance are identified as task-related brain areas. GLM has helped to identify the working memory-related brain regions~\cite{grady2021influence} and the Chinese language-related brain region~\cite{wu2013dissociable}. 
Recently, machine learning has been applied to uncover the brain regions and activity patterns across tasks and stimuli, such as multi-voxel pattern analysis (MVPA)~\cite{walther2016reliability} and graph convolutional networks~\cite{zhang2021functional}. Ye et al. proposed a Spatial Temporal-pyramid Graph Convolutional Network (STpGCN) to decode 23 cognitive tasks and a model-agnostic explanation tool (BrainNetX) to identify task-related brain regions~\cite{ye2023explainable}. 
Although these machine learning methods succeed in identifying task-relevant brain regions, they fall short in extracting the explicit time courses of inputs to the brain.
The ultimate goal is the reverse engineering of brain inputs, which involves uncovering the relationship between endogenous or exogenous stimuli and neural responses in complex cognitive tasks. 

Network control theory offers versatile tools for investigating the brain system. For instance, it can model input-output relationships in the brain with precision \cite{yang2021modelling} and elucidate the brain state transitions driven by minimal energy\cite{gu2017optimal,mccormick2020neuromodulation}.
Previous studies have shown that the brain network can be controlled by specific regions\cite{gu2015controllability}. This finding suggests that altering brain states does not require stimulation inputs to all nodes. Nevertheless, in practical scenarios, considering the large amount of energy input required, changing the brain states by input to a single brain node is unrealistic. It is widely recognized that multiple brain regions receive inputs during cognitive tasks. Adopting the terminology from control theory, we refer to these brain regions as \textit{control nodes}. Identifying control nodes and determining the input sequences of these nodes represent two key questions in this work.

Here, we present an input identification framework for the reverse engineering of brain inputs during cognitive tasks. We formulate this as a joint optimization problem: i) identifying critical control nodes and ii) inferring their corresponding control sequences. 
The objective function aims to minimize the reconstruction error of the neural dynamics using a sparse control set and smooth inputs. To solve this objective function, we propose a gradient-based augmented Lagrangian method, taking into account the consideration of the relaxation of boolean variables. 
The results from synthetic data validate the feasibility of inferring the control nodes and the input sequence. We apply our framework to the task fMRI data, identifying the motor-related brain regions and the explicit input sequence during motor tasks. 
Our framework sheds light on the underlying principles linking control nodes and the topological properties of the brain. 
\begin{figure}[t]
\centerline{\includegraphics[width=0.9\linewidth]{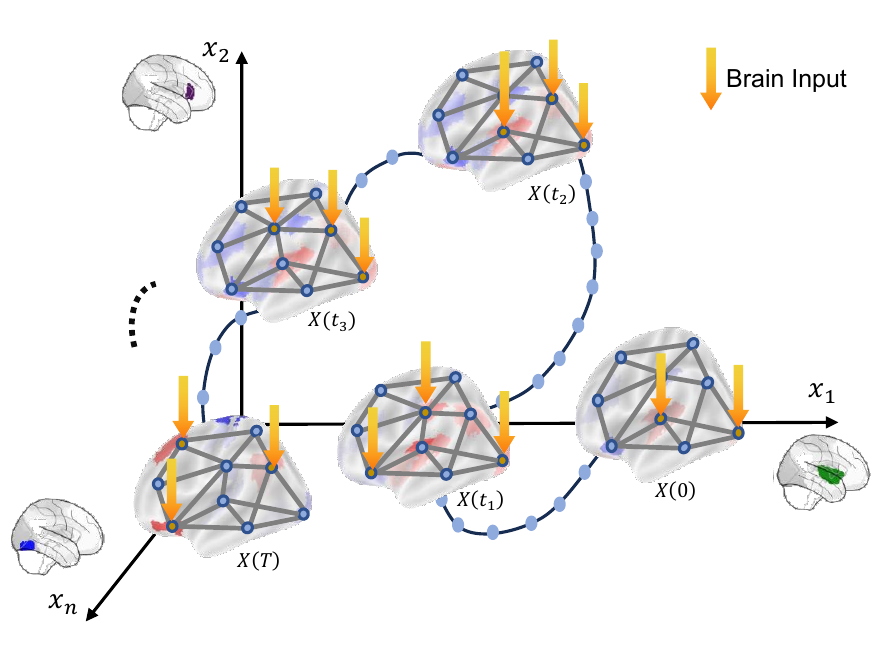}}
\caption{Inferring control nodes and input signals. The brain network system is modeled with a linear dynamic model
, $\mathbf{x}(t+1)=\mathbf{A}\mathbf{x}(t)+\mathbf{B}\mathbf{u}(t)$. The goal is to identify the control node $B$ and the input sequence $\mathbf{u}(t)$ given the observed fMRI dynamics $\mathbf{x}(t)$ and the pre-defined $\mathbf{A}$. 
}
\label{framework}
\end{figure}

\section{The input identification framework}

We present a control theory-driven framework designed to identify brain inputs during cognitive tasks. The framework consists of a dynamic model of the brain system, an objective function with associated constraints, and an optimization algorithm.
It is important to note that, although we implement a linear dynamic model in this study, this framework can be extended to any input-output models (e.g., SINDy~\cite{brunton2016sparse} and Koopman~\cite{liang2022online}).

\subsection{Brain network dynamical model}

Here, we model the brain system as a discrete-time linear time-invariant network with input (Fig.~\ref{framework}). The brain network model is formulated as Eq~\eqref{eq:system}.
\begin{equation} \label{eq:system}
    \mathbf{x}(t+1)=\mathbf{A}\mathbf{x}(t)+\mathbf{B}\mathbf{u}(t),
\end{equation}
where vector $\mathbf{x}(t) \in \mathbb{R}^N$ denotes the state of the brain network with $N$ regions at time $t$. Within our brain network model, the symmetric matrix  $\mathbf{A} \in \mathbb{R}^{N \times N}$ is the state transition matrix describing the interactions between brain nodes; the diagonal matrix $\mathbf{B} \in \mathbb{R}^{N \times N}$ is the input matrix with the value in diagonal either 1, indicating the brain region is selected as a control node, or 0, indicating the region is not a control node; the vector $\mathbf{u}(t) \in \mathbb{R}^N$ represents the control strategy to the brain network at time $t$.

In this study, our goal is to infer the input signals of the brain system during specific tasks according to the observed fMRI data. We define $\mathbf{A}$ as the structural connectivity of the brain regions, which can be obtained by DTI imaging. The system dynamics $\mathbf{x}(t)$ are observed, reflected in fMRI data. The parameters $\mathbf{B}$ and $\mathbf{u}(t)$ remain unknown.
Technically, the key challenge lies in identifying the binary input matrix $\mathbf{B}$—representing critical control nodes in the brain network. To this end, we propose a neural perturbation-based approach to identify $\mathbf{B}$, and employ a data-driven approach to infer the input signal $\mathbf{u}(t)$ from the task fMRI data. 


\subsection{Objective and optimization of input identification}
\subsubsection{Objective function and contraints}
Since we model the brain system with explicit expression defined by Eq.~\eqref{eq:system}, the input identification problem can be conceptualized as a parameter estimation problem by fitting data.
Inspired by machine learning and neuroscience prior knowledge, we define the objective function with three terms, including a \textit{data reconstruction loss} to minimize the reconstruction error and two \textit{input regularization} terms to ensure the smoothness of brain inputs. The brain system is constrained by the linear network dynamic model and input matrix $\mathbf{B}$. The objective function with system constraints is shown in Eq.~\eqref{eq:problem1},
\begin{equation} \label{eq:problem1}
 \begin{aligned}
 \min _{B, U_{1:T}} &\sum_{t=1}^T\|\mathbf{x}(t)-\widehat{\mathbf{x}}(t)\|_2^2+\lambda_1 \sum_{t=1}^T\|\mathbf{u}(t)\|_2^2 \\
 &+\lambda_2 \sum_{t=1}^{T-1}\|\mathbf{u}(t+1)-\mathbf{u}(t)\|_2^2\\
\text { s.t. :  } &\mathbf{x}(t+1)=\mathbf{A} \mathbf{x}(t)+\mathbf{B}\mathbf{u}(t) \\
& \sum_{i=1}^N \mathbf{B}_{i,i} =\alpha, \quad\quad \mathbf{B}_{i,i} \in\{0,1\},
\end{aligned}
\end{equation}
where $\mathbf{x}(t) \in \mathbb{R}^{N}$ is the state of the brain network at time step $t$; $\widehat{\mathbf{x}}(t)$ is the generative state in the brain network control framework; $\mathbf{A}$ is an $N \times N$ adjacency matrix that describes the structural connections among brain regions; $\mathbf{B}$ is the control matrix where $\mathbf{B}_{i,i} \in \{0, 1\}$ for $i=\{1,2,\dotsc, N\}$, indicating whether the node $i$ is a critical node in the specific task. There are in total three hyperparameters, \textit{i.e.,} $\lambda_1$, $\lambda_2$, and $\alpha$. We use the hyperparameters $\lambda_1$ and $\lambda_2$ to constrain the energy and the slope of input signals, respectively. The hyperparameter $\alpha$ determines the number of identified critical nodes.

\subsubsection{Optimization Algorithm}
Note that Eq~\eqref{eq:problem1} is a nonconvex optimization problem due to the Boolean constraints $\mathbf{B}_{i,i} \in \{0, 1\}$, making it a challenge to solve directly. Here, we relax the Boolean constraint using continuous epsilon relaxation with $|\mathbf{B}_{i,i} (1-\mathbf{B}_{i,i})| \leq \epsilon$,
which yields the following optimization problem,
\begin{equation} \label{eq:problem2}
 \begin{aligned}
 \min _{\mathbf{B}, \mathbf{u}_{1:T}} &\sum_{t=1}^T\|\mathbf{x}(t)-\widehat{\mathbf{x}}(t)\|_2^2+\lambda_1 \sum_{t=1}^T\|\mathbf{u}(t)\|_2^2 \\
 &+\lambda_2 \sum_{t=1}^{T-1}\|\mathbf{u}(t+1)-\mathbf{u}(t)\|_2^2\\
\text { s.t. :  } &\mathbf{x}(t+1)=\mathbf{A} \mathbf{x}(t)+\mathbf{B} u(t) \\
& \sum_{i=1}^N \mathbf{B}_{i,i} =\alpha, \quad |\mathbf{B}_{i,i} (1-\mathbf{B}_{i,i})| \leq \epsilon_k, 
\end{aligned}
\end{equation}
where $\epsilon_k$ is a violation parameter that smoothes the integer constraints and it will converge to a sufficiently small number during the optimization. 

To solve the optimization problem in Eq.~\eqref{eq:problem2}, 
we utilize the Augmented Lagrangian Method (ALM) with gradient descent for iterative optimization, progressively converging towards the optimal solution. 
Importantly, similar to training recurrent neural networks, the gradient of $\mathcal{L}$ to $B$ and $U$ can be computed automatically by the $AUTOGRAD$ in Pytorch and updated efficiently by backpropagation through time (BPTT).

\begin{figure}[t]
\centerline{\includegraphics[width=0.92\linewidth]{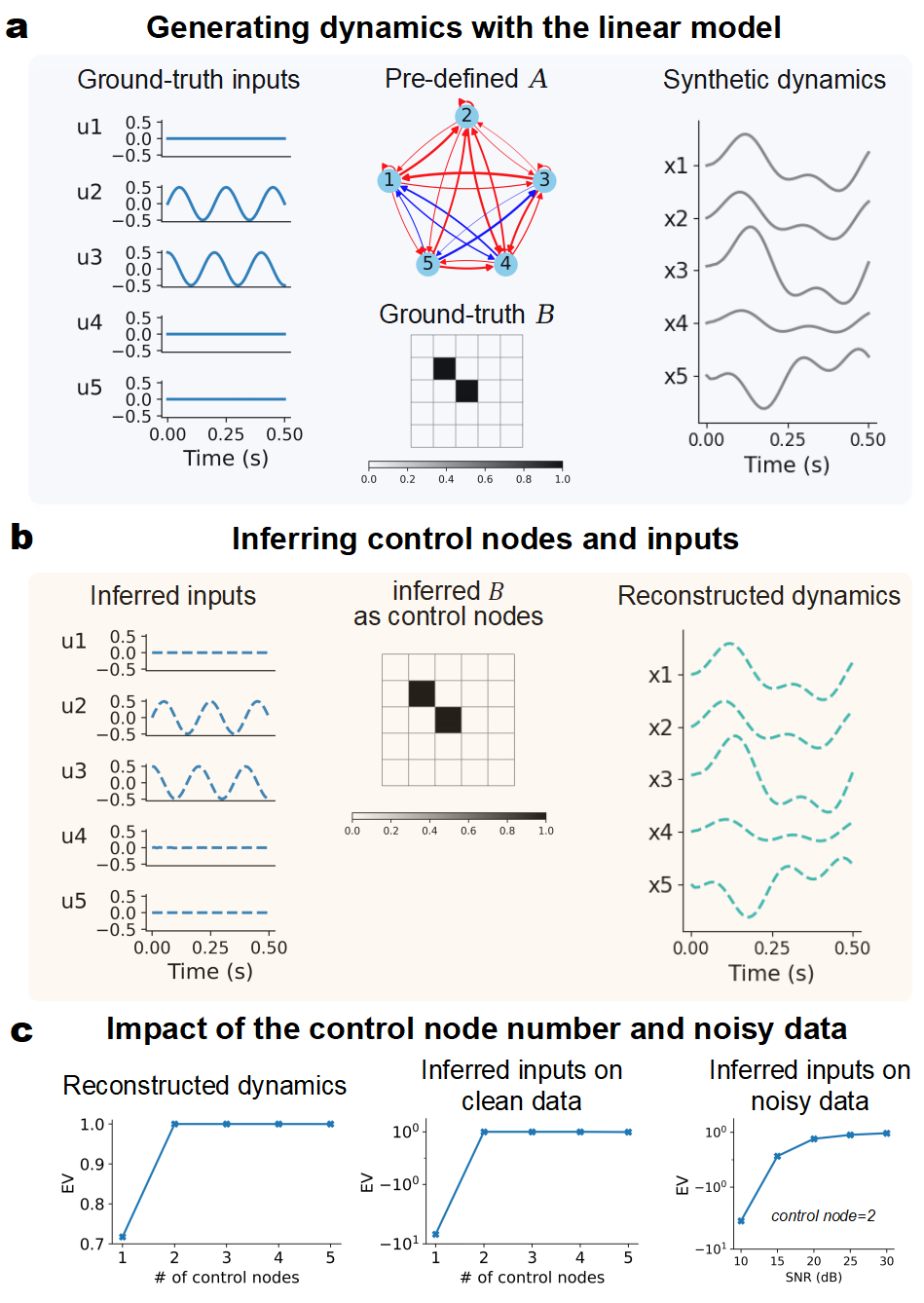}}
\caption{Simulating data and reverse engineering inputs. \textbf{a} The generated dynamics by a linear model. \textbf{b} Solution of the reverse problem, including the inferred control nodes (left), their corresponding inputs (middle) and the reconstructed dynamics (right). \textbf{c} The explained variance (EV) between the reconstructed and generated dynamics (left), between the inferred and ground-truth inputs on clean data (middle) and noisy data (right), respectively. The higher the EV, the better the performance.}
\label{fig:generate_dynamics}
\end{figure}

\section{Results}
We apply the framework to both synthetic data generated by a pre-defined linear dynamic model and the motor task-fMRI data to validate the efficacy of our approach.

\subsection{Validation with simulated data}
We first validate the performance of the proposed algorithm by addressing the reverse problem of deducing control nodes and their corresponding inputs from synthetic data generated by a pre-defined linear model. Specifically, we randomly initialize a state-transition matrix $\mathbf{A}$ with 6 nodes, and two input sequences $sin(10\pi t)$ and $cos(10\pi t)$ to the node 2 \& 3, respectively. Following the linear model in Eq.~\eqref{eq:system}, we synthesize the evolving dynamics (Fig.~\ref{fig:generate_dynamics}a). 
Given the synthetic data, we apply our method to iteratively update $\mathbf{B}$ and $\mathbf{u}$, and ultimately it converges to a steady solution (Fig.~\ref{fig:generate_dynamics}b). The quantitative results demonstrate the effectiveness of our method on both clean and noisy data (Fig.~\ref{fig:generate_dynamics}c).

\begin{figure}[t]
\centerline{\includegraphics[width=0.95\linewidth]{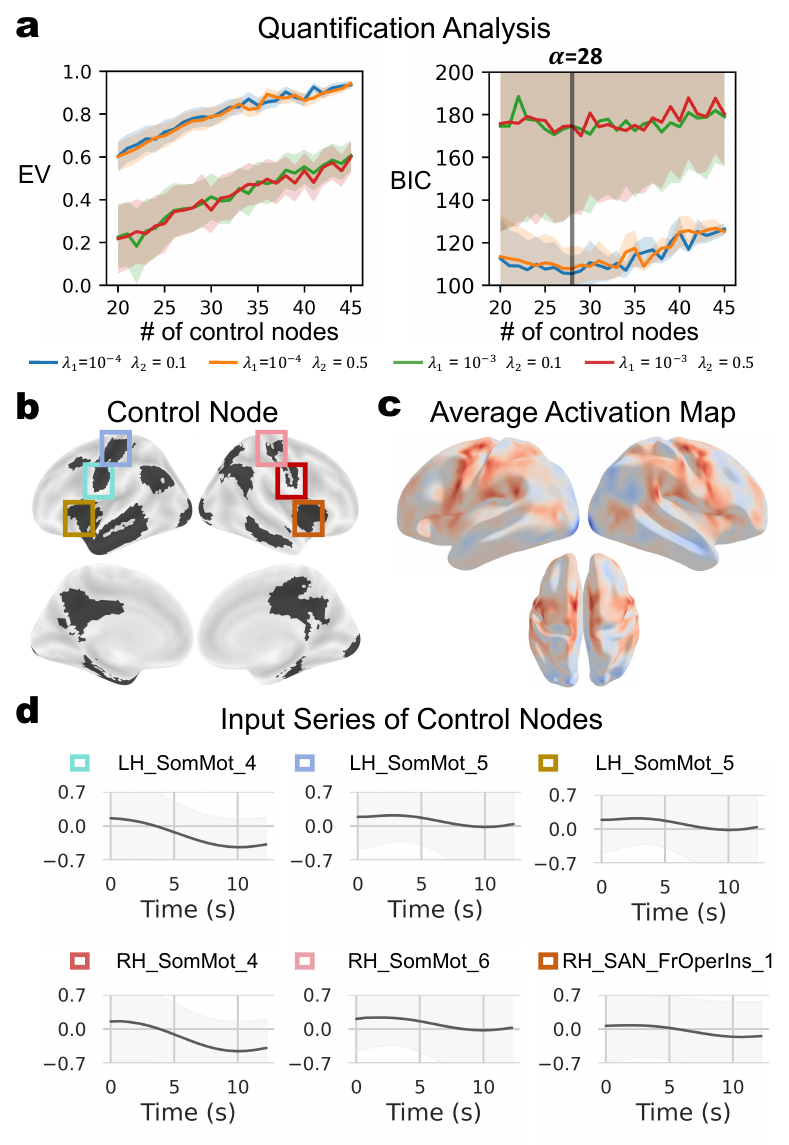}}
\caption{Motor task fMRI data and reverse engineering inputs. \textbf{a} Quantification analysis on tuning hyperparameters (the best hyperparameters set: $\lambda_1=0.0001,\lambda_2=0.1 \text{ and } \alpha=28$). \textbf{b} The visualization of the inferred control nodes. \textbf{c} The average activation map of the motor task in the HCP dataset. \textbf{d} Input series of ROIs in the identified control nodes.}
\label{fig:motor_task}
\end{figure}

\subsection{Identifying the inputs using fMRI data under motor task}

\subsubsection{Motor task and fMRI recording}
To validate the proposed framework, we utilized the task-fMRI dataset from the Human Connectome Project (HCP 1200 release). We randomly selected 200 out of 1200 subjects. All selected subjects include T1 MRI, dMRI, and fMRI recording during the motor tasks. The motor tasks comprised of 5 sub-tasks, including left/right foot, left/right hand, and tongue. Each motor task includes 17 Repetition Times (TRs) with a $TR=0.72$ second.
We parcellate the brain into 100 non-overlapping regions of interest (ROIs) using Yeo atlas\cite{yeo2011organization}. 
The $100\times100$ structural connectivity is obtained from DTI images, and $100 \times 17 \times trials$ time courses are extracted from motor task-fMRI data (trials = 200 subjects $\times$ 5 sub-tasks). 

\subsubsection{Identified control nodes and inputs}
We extend our framework to identify the critical control nodes with empirical task-fMRI data. For the group-level analysis, we first apply two metrics to quantify the performance of fitting accuracy across max control nodes and regions of the whole brain, including the Explained-Variance ($EV$), and the  Bayesian Information Criteria ($BIC$). $EV$ quantifies the fitting accuracy of the task-related functional dynamics of the whole brain across the number of control nodes (From 25 to 45) (Fig.~\ref{fig:motor_task}a). Fig.~\ref{fig:motor_task}a (top) shows the mean and variance of global whole-brain $EV$ across tasks with different numbers of control nodes in each subject. The curves increase with the increase of $\alpha$ until they reach a value with high predictive performance. Meanwhile, we apply $BIC$ to select the optimal number of control nodes. Fig.~\ref{fig:motor_task}a (bottom) shows the mean and variance of $BIC$ across tasks with different numbers of control nodes in each subject. It shows that a bell-shaped curve with the increase of $\alpha$, and the optimal number of control nodes is $\alpha=28$ of motor tasks. 

To interpret the identified nodes, we compare them with the average-level activation map of motor tasks in the HCP dataset (Fig.~\ref{fig:motor_task}b\&c). We found that most of the inferred control nodes are well in line with the activated map, largely overlapping in motor and somatosensory areas. The inferred input sequences of ROIs are illustrated in Fig.~\ref{fig:motor_task}d.

\section{Discussion and conclusion}

\subsubsection*{From methodology perspective}
Our study presented a general framework based on network control theory for identifying brain inputs during cognitive tasks. We proposed the Augmented Lagrangian Method with gradient descent to iteratively search the steady state of the objective function, which guarantees finding the optimal with the sparse control nodes and the smooth input sequences. The framework was validated with both synthetic data and real task-based fMRI data from the HCP dataset. 
Different from statistical analyses~\cite{grady2021influence} and machine learning methods~\cite{walther2016reliability}, we brought the network control theory to node identification. Rather than focusing on partially observed data, we use the whole-brain fMRI to infer the system parameters $\mathbf{B}$ and $\mathbf{u}(t)$.

\subsubsection*{From neuroscience perspective}
Revealing how external stimuli are processed by the brain and dynamically transmitted in different brain regions is the core question in neuroscience. Our work delves into the brain's input-output response to task stimuli, focusing on two key aspects: identifying control nodes and inferring input sequences.
Identifying control nodes enables us to pinpoint core brain regions relevant to cognitive functions. Also, it is important in clinical applications, such as selecting electrical stimulation sites for precise brain function modulation\cite{wang2023multi}.
Moreover, characterizing the input sequences is key to understanding how information flows between brain regions \cite{manjunatha2024controlling}. The temporal dynamics elucidate the information process across brain regions and the integration and coordination of information during complex cognitive tasks.
%


In summary, we proposed a new framework from control theory perspective for understanding brain inputs and system dynamics during cognitive tasks. It is a potential tool for unraveling the input-output relations in the brain under task stimulation and electrical stimulation.

\bibliographystyle{IEEEtran}
\bibliography{ref}
\end{document}